\documentclass[twocolumn,preprintnumbers,amsmath,amssymb,showpacs,prb,aps]{revtex4}
\usepackage{txfonts}
\usepackage{pifont}
\usepackage{amssymb}
\usepackage{dcolumn}
\usepackage{amsmath}
\usepackage[dvips]{epsfig}

\makeatletter
\def\btt#1{\texttt{\@backslashchar#1}}%
\DeclareRobustCommand\bblash{\btt{\@backslashchar}}%
\makeatother

\begin{document}
\title{Effect of Spin Correlations on Multi-orbital Metal-Insulator Transitions
    and Suppression of Orbital Selective Mott Transitions }
\author{Ya-Min Quan$^{1}$, Liang-Jian Zou$^{1,2 \footnote{Correspondence author,
        Electronic mail: zou@theory.issp.ac.cn}}$ and Hai-Qing Lin$^{2}$}
\affiliation{ \it $^1$ Key Laboratory of Materials Physics,
              Institute of Solid State Physics, Chinese Academy of Sciences,
               P. O. Box 1129, Hefei 230031, China\\
         \it $^2$ Department of Physics, Chinese University of Hong Kong,
                  Shatin, New Territory, Hong Kong, China\\}
\date{Nov 6, 2010}

\begin{abstract}
We present the influence of spin correlation on the metal-insulator
transitions (MIT) in two-orbital Hubbard models by the
Kotliar-Ruckenstein slave-boson approach. In the asymmetric
half-filling situation, the two orbits simultaneously transit from
conducting to insulating states with the increase of Coulomb
correlation, accompanied by a paramagnetic (PM)-antiferromagnetic
(AFM) transition. The orbital selective Mott transition found in the
PM condition is completely suppressed over a wide correlation range,
though it may exist in the systems away from half-filling. In the
insulating state, the system crosses over from a partially-polarized
spin-gapped phase in the intermediate correlation regime to an
almost fully-polarized Mott insulating phase in the strong
correlation regime. These results demonstrate that the spin
modulation to the quasiparticle spectra brings much rich and more
interesting MIT scenario in multi-orbital correlated systems.

\end{abstract}

\maketitle


Orbital selective Mott transition (OSMT), {\it i.e.}, with the
increase of Coulomb correlation, a two-orbital electronic system
transits from metal to a particular phase with one orbit exhibiting
metallic conduction and another orbit being insulating, is an
interesting topic \cite{Anisimov}. Theoretically it has been known
that such an unusual phase is robust in asymmetric two-orbital
Hubbard models \cite{Koga,Buenemann,Zou1}. The OSMT phase can be
driven by the asymmetric factors of two orbits, for example, the
level splitting between two orbits and the different bandwidths or
band degeneration. There exist great debates on whether some typical
compounds with metal-insulator transitions (MIT), such as
Sr$_{2-x}$Ca$_{x}$RuO$_{4}$ and V$_{2-x}$Cr$_{x}$O$_{3}$, are the
prototypes of the OSMT phase \cite{Ding,Liebsch,Zou2}. This
naturally arises a question: whether does the OSMT phase really
exist in realistic compounds, especially in the magnetic ordered
systems?
Generally speaking, the Mott-Hubbard MIT are accompanied with the
change of magnetic structures. Such a change would crucially
modulate the quasiparticle states, hence significantly affect the
MIT. The MIT and OSMT phase have been studied widely based on the
dynamical mean-field theory (DMFT) \cite{Kotliar1} and
Kotliar-Ruckenstein slave boson (KRSB) methods \cite{Kotliar2}, both
of which are effective many-body approaches to treat the electronic
interaction over a wide correlation strength. However, limited by
the huge computation time in the DMFT and KRSB methods, most of the
studies focus on the paramagnetic (PM) phases. How the spin degree
of freedom affects the MIT phase diagram and whether the OSMT phase
exists in magnetic phase diagram are still not clear.

Hesagawa \cite{Hesagawa} and Sigrist {\it et al.} \cite{Sigrist}
first extended the single-orbital KRSB theory to the two-orbital
situation to study the PM Mott transitions in two-orbital correlated
electron systems. To explore the influence of the spin degree of
freedom on the Mott transitions in multi-orbital Hubbard models, we
apply the extended KRSB method \cite{Kotliar1} on asymmetric
two-orbital Hubbard models for various electron fillings. With the
help of a new numerical ansatz to the multi-orbital KRSB solution we
developed recently \cite{Quan}, which overcomes the convergency
problem of many parameters in minimizing the groundstate energy, we
could treat arbitrary interaction strengths, finite interorbital
hoppings or hybridizations, and various magnetic configurations in
multi-orbital correlated electron systems.
In this Letter, starting with an asymmetric two-orbital Hubbard
model, we present various magnetic phase diagrams of the MIT, and
discuss the physical scenario of the MIT in the presence of spin
degree of freedom. It is interestingly found that the OSMT phase
completely vanishes at half filling. A Slater-type spin-gapped
insulator and an orbital polarized band insulator are also found in
magnetic phase diagrams, in addition to the conventional PM and {\it
N\'{e}el} antiferromagnetic (AFM) metallic phases. Away from half
filling, an AFM OSMT phase may be stable. These results show that
the magnetic MIT picture in multi-orbital correlated systems is more
interesting and much richer than that without spin.

\begin{figure}[htbp]
\centering
\includegraphics[angle=0, width=0.9 \columnwidth]{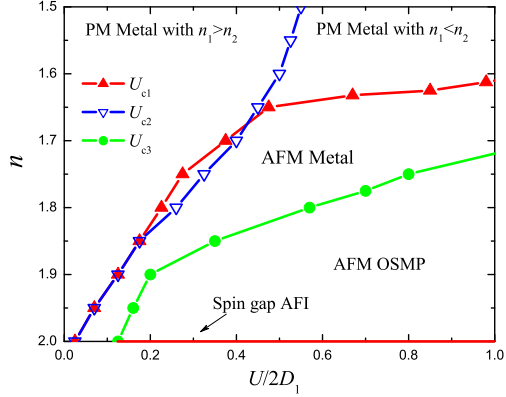}
\caption{Magnetic phase diagram of the two-orbital Hubbard model in
the $n-U$ plane. $U_{c1}$ and $U_{c3}$ denote the
PM metal-AFM metal and the AFM metal-AFM insulator boundaries.
$U_{c2}$ separates different orbital polarization regions.
Theoretical parameters: $D_{2}/D_{1}=0.5$, $J_{H}=0.2U$ and
$\Delta=0$.} \label{fig:1}
\end{figure}

Our starting point is an asymmetric two-orbital Hubbard model
\cite{Hesagawa,Zou1} applicable for various correlated electron
systems
\begin{eqnarray}
  H&=&H_{0}+H_{I}\\
  H_{0}&=&-\sum_{ij\alpha\beta\sigma}\left(t_{\alpha\beta}c^{\dagger}_{i\alpha
  \sigma}c_{j\beta\sigma}+h.c.\right)+ \sum_{i\alpha\sigma}\left(\varepsilon_{
  \alpha}-\mu\right)n_{i\alpha\sigma}  \\
  H_{i}&=&U\sum_{i\alpha}n_{i\alpha\uparrow}n_{i\alpha\downarrow}+\sum^{\left(
  \alpha>\beta\right)}_{i\sigma\sigma^{\prime}}
   \left(U^{\prime}-J_{H}\delta_{\sigma\sigma^{\prime}}\right)n_{i\alpha\sigma}
   n_{i\beta\sigma^{\prime}}               \nonumber\\
   & &-J_{H}\sum_{i\alpha\neq\beta}\left(c^{\dagger}_{i\alpha\uparrow}c_{i\alpha
   \downarrow}c^{\dagger}_{i\beta\downarrow}c_{i\beta\uparrow}
   -c^{\dagger}_{i\alpha\uparrow}c^{\dagger}_{i\alpha\downarrow}c_{i\beta
   \downarrow}c_{i\beta\uparrow}\right),
\end{eqnarray}
on a square lattice with only the nearest-neighbor hopping
integrals. Where $t_{\alpha\beta}$, $U$ ($U^{\prime}$) and $J_{H}$
represent the hopping integral for the orbits $\alpha$ and $\beta$,
the intraband (inter-band) Coulomb repulsion and Hund's rule
coupling, respectively. Throughout this paper we set $U^{\prime}$=
$U-2J_{H}$, and take the energy bandwidth of the first orbit,
$2D_{1}=2zt_{11}=2zt$, be the energy unit.
To explore the effect of spin in the multiorbital MIT, we introduce
a few of auxiliary boson field operators representing the
possibility of various electron occupations, such as $e$,
$p_{\alpha\sigma}$, $d_{\sigma_{\alpha}\sigma'_{\beta}}$,
$b_{\alpha\beta}$, $t_{\alpha\sigma}$ and $q$, which denote various
possibilities of none, single, double, triplicate and quadruplicate
electron occupations.
%
%
Projecting the original fermion operators into these boson field and
fermion field operators, one can obtain an effective Hamiltonian,
from which one can get the groundstate energy in the saddle point
approximation. We apply the generalized Lagrange multiplier method
to enforce the fermion number constraint condition \cite{Hesagawa}.
In this framework the interorbital hoppings and crystal field
splitting can be treated in the same foot.
%
%

We adopt an optimizing method to get the minimized groundstate
energy for four different magnetic configurations, including the PM
or nonmagnetic, ferromagnetic (FM) and {\it N\'{e}el} AFM states.
%
%
In the optimizing process the boson field normalization set a
boundary for the mean value of each boson field. As shown in our
previous work \cite{Zou3}, the optimizing problem of boundary
constrained condition is still difficult. In numerically searching
the global minima of the groundstate energy, we combine the pattern
search method, the gradient method and the Rosenbrock method. As the
optimizing point is approaching to the boundary, one must move one
step inward the high-dimensional ellipsoid and the equipotential
plane. Since the first axis of the new local orthogonal coordinate
system of Rosenbrock method direct to the negative gradient
direction, such an algorithm can be easily achieved. Our ansatz
obtains a lot of desirable results for various PM systems with
single, two and three orbitals \cite{Quan}


\begin{figure}[htbp]
\centering
\includegraphics[angle=0, width=0.7 \columnwidth]{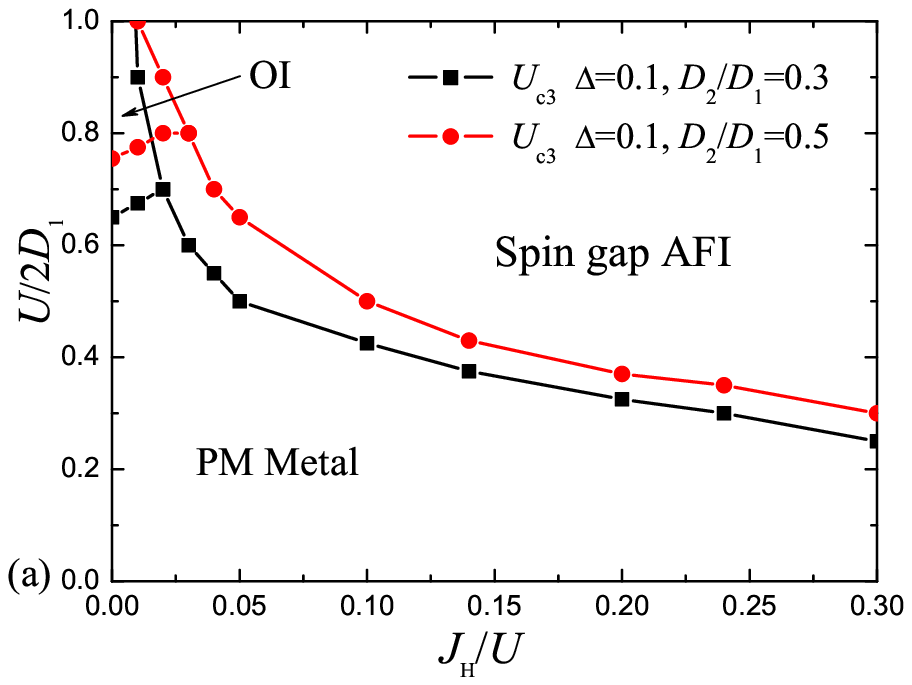}
\includegraphics[angle=0, width=0.7 \columnwidth]{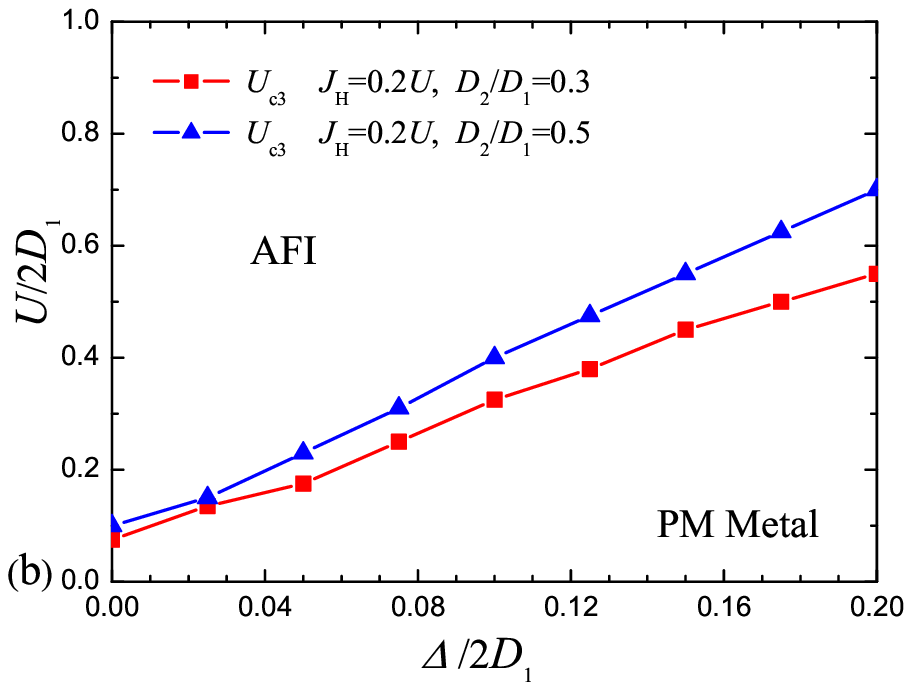}
\caption{Magnetic phase diagram of half-filled two-orbital Hubbard
model in the U-J$_{H}$ plane with $\Delta/2D_{1}=0.1$ (a) and in the
U-$\Delta$ plane with J$_{H}$=0.2U (b) for the bandwidth ratio
D$_{2}$/D$_{1}$=0.3 and 0.5. OI and AFI denote the orbital insulator
and the AFM insulator, respectively.} \label{fig:2}
\end{figure}

Fig.\ref{fig:1} is magnetic phase diagrams of the asymmetric
two-orbital Hubbard model with the bandwidth ratio
$D_{2}/D_{1}=0.5$, the Hund's rule coupling $J_{H}=0.2U$ and the
crystal field splitting $\Delta=\varepsilon_{2}-\varepsilon_{1}=0$.
We only plot the phase boundaries for the electron filling of $n
\leq 2$, the magnetic phase diagram for $n>2$ can be mapped by
considering the particle-hole symmetry character in the present
square lattice with the nearest-neighbor hopping. On the whole, we
find four different stable magnetic phases in Fig.1, including a PM
metal, an AFM metal, an AFM OSMT phase and an AFM insulator. The
boundaries between these different phases are U$_{c1}$ and U$_{c3}$.
Out of our expectation, the PM OSMT phase is not found for all of
the electron filling and correlation strength. We address this
magnetic phase diagram in length in the following.
%
%

Firstly, we find that in the half filling (n=2) and $\Delta$=0, with
the increase of the electron correlation the system undergoes
transitions from a PM metal (only for U=0) to an AFM metal at the
the PM metal-AFM metal transition boundary U$_{c1}$, and then to an
AFM insulator at the AFM MIT boundary U$_{c3}$, as seen in Fig.1. It
is obviously that the critical line $U_{c3}$ is considerable smaller
than that without the spin degree of freedom \cite{Jakobi}. Contrary
to the early and recent results without the spin correlation
\cite{koga,Jakobi}, the OSMT phase is completely suppressed in the
present situation. Since the spin degree of freedom is taken into
account, the present AFM insulator is a spin gapped state in the
intermediate correlation region, or a AFM Mott state in the large U
region. Approximately, the spin gapped insulating region at half
filling covers the PM OSMT phase region obtained by the DMFT
approach \cite{Jakobi}. This shows that when U increases, the narrow
orbit of the system becomes of AFM insulating. Such an AFM
correlation also modifies the electronic properties in the wide
band, and leads the spectra of the wide-band electron to open a gap.
Hence the MIT simultaneously occurs in the two-orbital system. As a
result, the PM OSMT phase is completely suppressed at half filling.

Away from half filling, the magnetic phase diagram becomes of more
interesting. As seen in Fig.1, though the PM OSMT is completely
suppressed, the AFM OSMT phase can survive over a wide region, in
which the half-filled narrow band becomes of AFM and insulating,
while the wide band away from half filling remains conducting,
though modulated by the AFM correlation.
We find that with the increase of the Coulomb correlation, the
system undergoes a PM metal-AFM metal transition and an AFM
metal-AFM OSMT transition. These transitions are also separated by
the critical line U$_{c1}$ and U$_{c3}$, respectively. Even in the
PM phase, the system may show different orbital polarization
separated by the critical line U$_{c2}$, as indicated in Fig.1. When
$U<U_{c2}$, the particles are mainly distributed in the wide band
and the reverse particle distributions achieve when $U>U_{c2}$,
since the Coulomb interaction drives a charge redistribution with
increasing $U$ \cite{Sigrist,Jakobi}. We notice that the critical
line U$_{c2}$ in Fig. \ref{fig:1} is almost independent of the
magnetic transition, suggesting that the charge redistribution in
different orbits is directly driven by the electron correlation,
rather than the spin exchange splitting.

Though the PM OSMP is suppressed in the presence of magnetic
correlation at half filling and $\Delta=0$, it is unknown whether it
exists over wide crystal field splitting $\Delta$ and the Hund's
rule coupling. In what follows, we focus on these situations in the
half-filled Hubbard model.
The magnetic phase diagrams for the half-filled asymmetric Hubbard
models in the $U-J_{H}$ and $U-\Delta$ planes are plotted in
Fig.\ref{fig:2}. It is still
%
%
find that the OSMT phase is absent in Fig.2a and Fig.2b. Also, as
seen in these two figures, the AFM metallic phase is unstable for
finite crystal field splitting, it will become of a PM metallic one.
Thus increasing the crystal field splitting may drive a AFM-PM
transition.

Fig.\ref{fig:2}a shows the $U-J_{H}$ phase diagram for different
bandwidth ratios.
%
%
In the PM metallic situation below the phase boundary U$_{c3}$, the
electrons in two orbits are itinerant and equivalently occupy two
orbits. The two orbits are singly occupied by two electrons and
become insulating and the PM-AFM MIT occurs as $U>U_{c3}$. In the
presence of spin degree of freedom, the system enters the
spin-gapped AFM insulating phase in the intermediate region in
Fig.2a. Due to the spin modulation, the PM-AFM MIT boundary $U_{c3}$
is considerable smaller than that without spin.
It is interesting that when the two orbits are split by a crystal
field, a new insulating phase, named orbital insulator, appears in
the small $J_{H}$ region. In the orbital insulating phase, the two
electrons fully occupy the lower narrow orbits and the system is
nonmagnetic and enters into an insulating phase. Such an interesting
phase has also been obtained recently by Yu and Si \cite{Rong}. Due
to the neglect of the spin effect in Yu and Si¡¯s work, their
$U_{c}/2D_{1}$ is larger than unity, considerably larger than our
result, about $0.68$, in the system with $J_{H}=0$,
$D_{2}/D_{1}=0.3$ and $\Delta=0.1$.
With the further increase of the electronic correlation, one may
expect that the system crosses overs from a spin-gapped insulator to
a conventional Mott insulator, as we will show in Fig.3 later. These
unusual results show that the spin correlation has profound
influence on the quantum phases and the boarder of the MIT in
correlated electron systems.

%
%

On the other hand, the magnetic phase diagram as the function of $U$
and $\Delta$ is shown in Fig.\ref{fig:2}b. One can see that the OSMT
phase is also excluded. From the U-$\Delta$ phase diagram in
Fig.\ref{fig:2}b, we see that the critical line $U_{c3}$ of the
PM-AFM MIT decreases with the decrease of the bandwidth ratios
$D_{2}/D_{1}$. At the half filling, large Coulomb
correlation and Hund's rule coupling, which contribute the spin
exchange splitting, are needed to overcome the energy increase
arising from the crystal field splitting. So the MIT boundary
$U_{c3}$ lifts with the increase of $\Delta$.
It is interesting that the spin-gapped AFM insulator in Fig.1 and
Fig.2 is not observed in the previous literature. In this phase the
two orbits are not fully spin polarized, this allows spins to form
density wave order. Considering the fact that this insulator is
stable in the intermediate correlated regime, we attribute it to the
{\it Slater}-type insulator \cite{Gebhard} with a partial spin
polarization.

\begin{figure}[htbp]
\centering
\includegraphics[angle=0, width=0.48 \columnwidth]{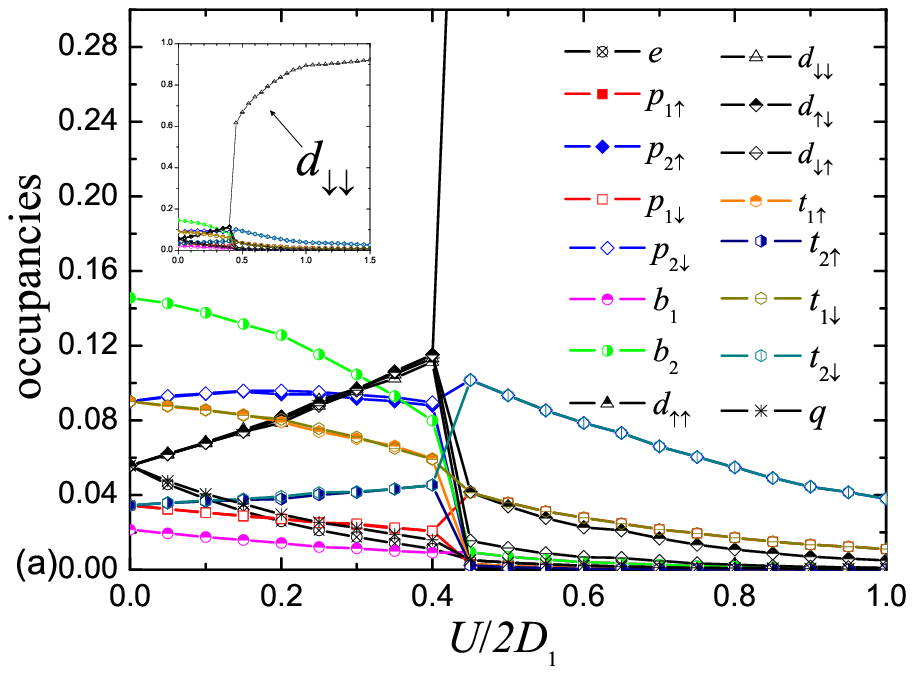}
\includegraphics[angle=0, width=0.48 \columnwidth]{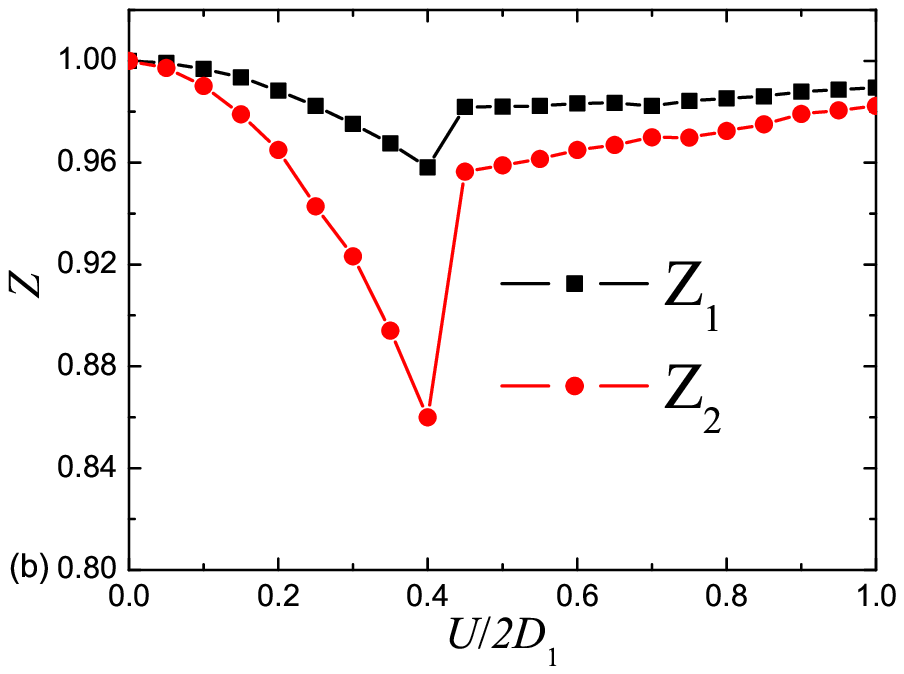}
\includegraphics[angle=0, width=0.48 \columnwidth]{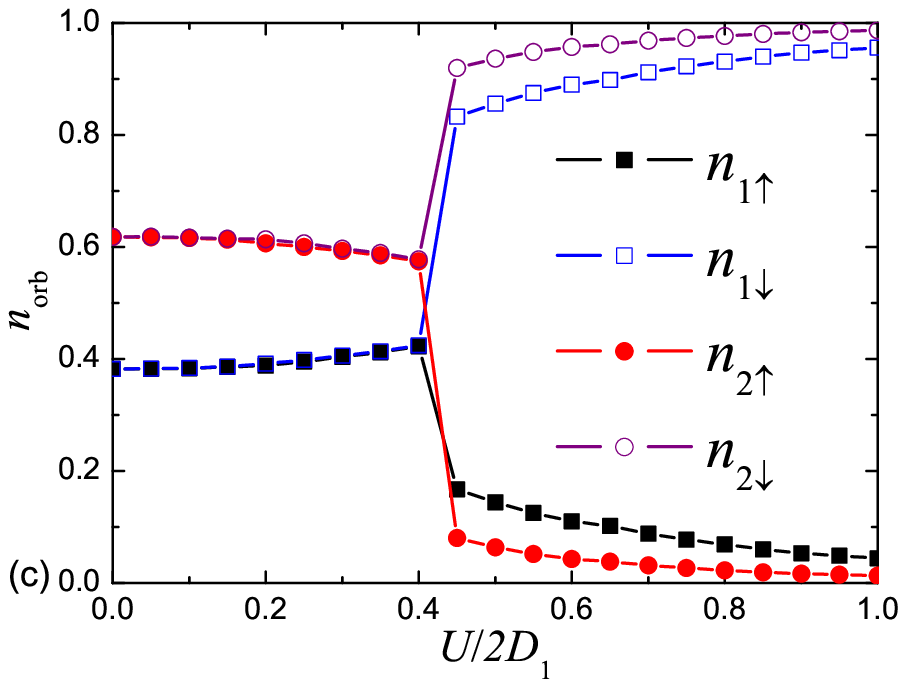}
\includegraphics[angle=0, width=0.48 \columnwidth]{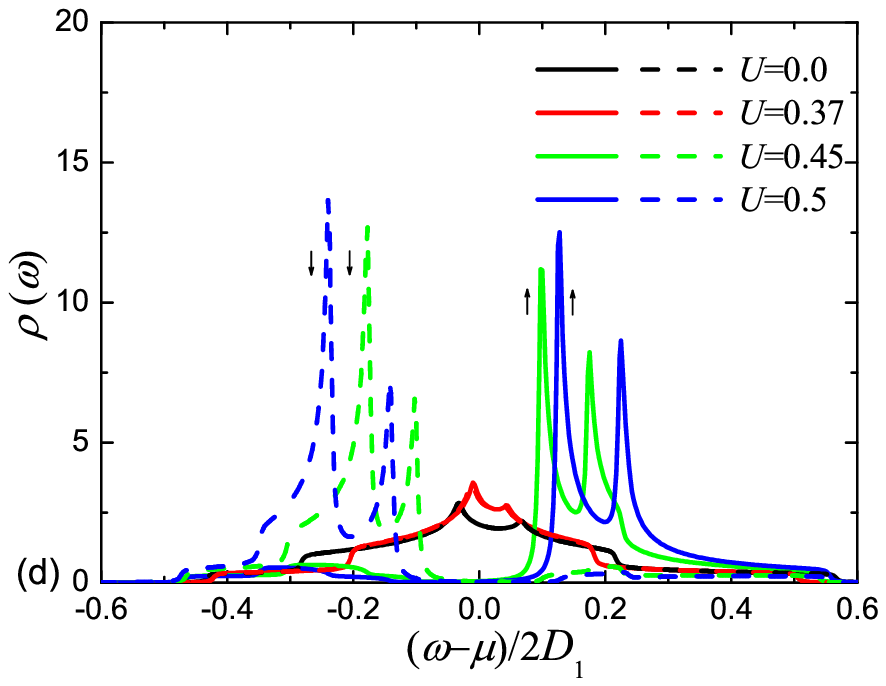}
\caption{Correlation dependence of occupation probabilities (a),
band renormalization factor (b), particle number (c) and density of
states (DOS) (d) of the half-filling two-orbital Hubbard model with
$D_{2}/D_{1}=0.5$, $J_{H}=0.2U$ and $\Delta/2D_{1}=0.1$. Inset in
(a) displays the full data of the boson occupation probability.}
\label{fig:3}
\end{figure}

Fig.\ref{fig:3} shows the correlation dependence of occupation
probabilities, band renormalization factor, particle number and DOS
for finite Hund's rule coupling in the half-filled two-orbital
Hubbard model. All these figures clearly show that the MIT occurs at
the critical point $U_{c3}$, accompanied by a magnetic transition.
Fig.\ref{fig:3}a displays the $U$ dependence of various occupation
probabilities. It shows that a first-order-type MIT transition
occurs at $U_{c3}=0.42$: when $U<U_{c3}$, various charge and spin
configurations appear in approximately equal weight, suggesting
that the PM metallic ground state is stable. When $U>U_{c3}$, the
spin-triplet $d_{\sigma\sigma}$ occupation is dominant, the system
enters the magnetically ordered insulating phase. As seen from
Fig.\ref{fig:3}a, only the $d_{\downarrow\downarrow}$
occupancy with each orbit occupied by one electron increases with the lift of
$U$ and $J_{H}$, the other occupancies decrease immediately when
$U>U_{c3}$. After $U>=2D_{1}$, all of the occupancy probabilities almost approach
constants. This indicates that the system enters a conventional
Mott insulating phase.

Fig.\ref{fig:3}b shows the dependence of the band renormalization
factor $Z_{\alpha}$ of the quasiparticles in orbit-$\alpha$ on the
Coulomb correlation $U$,
which significantly differs from the behavior of $Z$ without
considering the spin correlation effect. In accordance with Fig.\ref{fig:3}a, we
find that with the increase of $U$, the renormalization factor
decreases in the PM phase since the electronic correlation narrows
the bandwidthes. Accompanied with the
transition from the PM metallic phase to the AFM insulating one, an
insulating spin gap appears.
Out of our expectation, it abruptly increases at the MIT critical
point $U_{c3}/2D_{1}=0.42$, and then gradually increases abnormally. With
the increase of $U/2D_{1}$ to about $1.0$, both $Z_{1}$ and $Z_{2}$
approach saturations. The system may cross over from a Slater-type
insulator to a conventional Mott-Heisenberg insulator, though both
of the band renormalization factors do not show such a crossover.
Our results are distinctly different from the PM results by Hesagawa
\cite{Hesagawa} and Sigrist {\it et al.} \cite{Sigrist}, where there
exists a stable and robust OSMT phase in the intermediate
correlation regime.

The $U$ dependence of the particle number of each orbit per
spin channel is shown in Fig.\ref{fig:3}c. It is clearly seen that
with the increase of the electronic correlation, the particle
numbers in the two orbits differ from each other
in the PM region and experience a critical change at $U_{c3}$, and
gradually approach constant occupations as $U/2D_{1}$ approaches unity,
in accordance with Fig.3a and Fig.3b.
Obviously the PM-AFM MIT at $U_{c3}$ is first order.
Although the spins are not polarized in the PM metallic phase, the
sublattice spins are strongly polarized in the insulating phase
($U>U_{c3}$) and the sublattice spin polarization almost saturates
as $U/2D_{1} \rightarrow 1$. Asymmetric bandwidths
and crystal field splitting lead the spin polarizations of two orbits
be different. The more interesting is that with the
transition from the PM to AFM phases when $U>U_{c3}$, the increase
of the Coulomb interaction depresses the orbital polarization, which
originates from the increase of the Hund's rule coupling.
%
%
We find that the finite crystal field splitting is the key reason for the
first-order MIT. When the crystal field splitting equals to zero,
the transition from the PM to magnetically ordered phases is the second order,
in agreement with Hasegawa's results \cite{Hesagawa}.

    The spin-dependent densities of states (DOS) for $U/2D_{1}=0, 0.37, 0.45$
and $0.5$ are plotted in Fig.\ref{fig:3}d. It demonstrates how the
two-orbital DOS evolve with the electronic correlation. In the PM
metallic phase, both bandwidths of the two orbits shrink with
increasing $U$. In the present square lattice, one finds two cusps
%
%
in the spin-dependent DOS of the two orbits when the crystal field
splitting is switched on. The MIT energy gap, which becomes large
with the increase of $U$, opens at $U_{c3}=0.42$ in the system with
D$_{2}$/D$_{1}$=0.5, J$_{H}$=0.2U and $\Delta/2D_{1}$=0.1. The
well-known quasiparticle peaks in Fermi surface and the three-peaks
structure found in the DMFT approach are not observed in the
spin-dependent multi-orbital KRSB theory, which is attributed to the
neglect of dynamical transition process between Hubbard subbands in
the present mean-field approach.


    From the preceding studies on the metal-insulator transitions and the
orbital selective Mott phase in the two-orbital Hubbard model with
spin degree of freedom in the slave boson mean field theory, we find
that in the presence of spin correlation, the scenario of the
metal-insulator transitions may be distinctly different from that in
the paramagnetic situations.
In the half-filled asymmetric two-orbital Hubbard models, the spin
modulation arising from the insulating narrow band opens a gap in
the wide band, leading the paramagnetic orbital selective Mott phase
to be suppressed over a wide correlation range; and the orbital
insulating phase is also found to stable in the small $J_{H}$
region. In the presence of spin correlations and finite crystal
field splitting, the metal-insulator transitions, accompanied by
magnetic transitions, are of first order.
Away from half-filling an antiferromagnetic orbital selective Mott
phase may survive and the critical point of U$_{c3}$ is smaller than
the PM results because the symmetry of the local interaction is
lowered by the spin polarization.
These results demonstrate that the spin correlations and the crystal
field splitting play very important roles for the metal-insulator
transitions.


This work was supported by the NSFC of China, the HKSAR RGC 401806
and the Knowledge Innovation Program of the Chinese Academy of
Sciences. Numerical calculations were performed at the Center for
Computational Science of CASHIPS.


\begin{thebibliography}{}

\bibitem{Anisimov}
 V. I. Anisimov, I.A. Nekrasov, D.E. Kondakov, T.M. Rice and M.
 Sigrist, {\it Euro. Phys. J.} {\bf B 25}, 191 (2002).
\bibitem{Koga}
 A. Koga, N. Kawakami, T.M. Rice and M. Sigrist, {\it Phys. Rev.
 Lett.} {\bf 92}, 216402 (2004); A. Koga, N. Kawakami, T.M. Rice and
 M. Sigrist, {\it Phys. Rev.} {\bf B 72}, 045128 (2005); A. Koga, K.
 Inaba and N. Kawakami, {\it Prog. Theo. Phys. Suppl.} {\bf 160}, 253
 (2005); K. Inaba and A. Koga, {\it Phys. Rev.} {\bf B 73}, 155106
 (2006).
\bibitem{Buenemann}
 J. Buenemann, D. Rasch and F. Gebhardet, {\it J. Phys.} Cond. Matt.
 {\bf 19}, 436206 (2007).
\bibitem{Zou1}
 Y. Song and Liang-Jian Zou, {\it Phys. Rev.} {\bf B 72}, 085114
 (2005).
\bibitem{Ding}
 M. Neupane, P. Richard, Z.-H. Pan, Y. Xu, R. Jin, D. Mandrus, X.
 Dai, Z. Fang, Z. Wang and H. Ding, {\it Phys. Rev. Lett.} {\bf 103}
 097001 (2009).
\bibitem{Liebsch}
 A. Liebsch and H. Ishida, {\it Phys. Rev. Lett.} {\bf 98}, 216403
 (2007).
\bibitem{Zou2}
 Y. Song and Liang-Jian Zou, {\it Euro. Phys. J.} {\bf B 72}, 59
 (2009).
\bibitem{Kotliar1}
 A. Georges, G. Kotliar, W. Krauth and M. J. Rosenberg, {\it Phys.
 Rev. Lett.} {\bf 57}, 1362 (1986).
\bibitem{Kotliar2}
 G. Kotliar and A. Ruckenstein, {\it Phys. Rev. Lett.} {\bf 57}, 1362
 (1986).
\bibitem{Hesagawa}
 H. Hasegawa, {\it J. Phys. Soc. Jpn.} {\bf 66}, 1391 (1997);
 {\it Phys. Rev.} {\bf B 56}, 1196 (1997).
\bibitem{Sigrist}
 A. R\"{u}egg, M. Indergand, S. Pilgram and M. Sigrist, {\it Eur.
 Phys. J.}, {\bf B 48} 55 (2004).
\bibitem{Quan}
 Y. M. Quan, Dayong Liu and Liang-Jian Zou, to appear in {\it Acta
 Physica Sinica} {2011} (Chinese).
\bibitem{Zou3}
 Y. M. Quan, Liang-Jian Zou, Dayong Liu and H. Q. Lin,
 arXiv:1104.4599.
\bibitem{Jakobi}
 E. Jakobi, Nils Bl\"{u}mer and Peter van Dongen, Phys. Rev. B {\bf
 80}, 115109 (2009).
\bibitem{Rong}
 Rong Yu and Qimiao Si , {arXiv:1006.2337v2}.
\bibitem{Gebhard}
 F. Gebhard, {\it The Metal-Insulator Transition: Models and
 Methods}, Spring, (1997).
%

\end{thebibliography}

\end{document}